\documentstyle[preprint,aps,epsf]{revtex}

\newcommand{\be}[1]{\begin{equation}\label{#1}}
\newcommand{\ee}{\end{equation}}
\newcommand{\bea}[1]{\begin{eqnarray}\label{#1}}
\newcommand{\eea}{\end{eqnarray}}
\newcommand{\bfi}[1]{\mbox{\boldmath ${{#1}}$}}
\newcommand{\phitil}{\tilde \phi}

\newcommand{\Jtil}{\tilde J}
\newcommand{\R}{\rm R}
\newcommand{\SR}{\rm SR}

\newcommand{\gsim}{\mbox{\hbox{ \lower-.6ex\hbox{$>$}\kern-.8em
\lower.5ex\hbox{$\sim$}\kern+.35em}}}

\begin{document}

\draft

\preprint{arch-ive/9506037}

\title{Crystal surfaces with correlated disorder: \\ Phase transitions
  between roughening and superroughening}

\author{Stefan Scheidl}

\address{Institut f\"ur Theoretische Physik, Universit\"at zu K\"oln,
  Z\"ulpicher Str. 77, D-50937 K\"oln, Germany}

\date{May 30, 1995}
\maketitle

\begin{abstract}
  A theory for surface transitions in the presence of a disordered
  pinning potential is presented. Arbitrary disorder correlations are
  treated in the framework of a dynamical functional renormalization
  group. The roughening transition, where surface roughness and
  mobility behave discontinuously, is shown to turn smoothly into the
  continuous superroughening transition, when the range of disorder
  correlations is decreased. Implications for random-field $XY$-models
  and vortex glasses are discussed.
\end{abstract}



The shape of a fluctuating crystal surface is affected by the pinning
potential provided by the crystal planes. The interface exhibits a
roughening transition, if these planes are perfect, and it exhibits a
superroughening transition, when these planes are strongly disordered.
The roughening transition is well understood theoretically and
verified in different experimental situations (see e.g.
\cite{BN77,NG87} and references therein). Above the roughening
temperature $T_{\R}$ the interface is thermally rough, whereas below
$T_{\R}$ the interface becomes smooth, since it locks into the pinning
potential. However, for quenched bulk disorder \cite{TdV90} and
substrate disorder \cite{RG,TS92} it was shown, that the transition
has a different nature. This so-called superroughening transition
occurs at $T_{\SR}=T_{\R}/2$. Above $T_{\SR}$ the interface is
thermally rough again, but below $T_{\SR}$ the disordered pinning
potential increases the roughness of the interface. In this Letter we
address the question: how are the roughening transition and the
superroughening transition related?  What happens ``in-between'', if
we switch from a pinning potential with long-range correlations to
short-range correlations?

As shown below, new physics emerges for logarithmic substrate
roughness. This can be realized, if the crystal is grown on a
logarithmically rough substrate. Such a substrate could be generated
by quenching it from a temperature above its roughening transition. In
order to have a stable substrate surface, the roughening temperature
of the substrate should be higher than that of the crystal.

Our analysis is also relevant for a $XY$-model in the presence of a
random field, as long as vortices can be neglected\cite{note.XY}.
Different types of sources of the random field may have different
multipolar character, which lead to a power-law decay of correlations.

Finally, the consideration applies to two-dimensional vortex glasses
\cite{vg,{HNV93}}. Their pinning can be caused by local suppressions
of the condensate density, which also relaxes with power laws and
leads in turn to a power-law decay of correlations. This power
varies with the dimensionality of the defects.

At present there is a controversial discussion about the actual
roughness of the superrough phase. Renormalization group (RG)
calculations \cite{RG,TS92}, variational calculations \cite{var} and
simulations \cite{simul} give an inconsistent picture. In view of this
controversy we focus on topics consistently described by these
different approaches: the phase boundary and the nature of the phase
transition.  The present study is based on a functional dynamical
renormalization group applying to the limit of very weak pinning.

\begin{figure}
\epsfxsize=\linewidth
\epsfbox{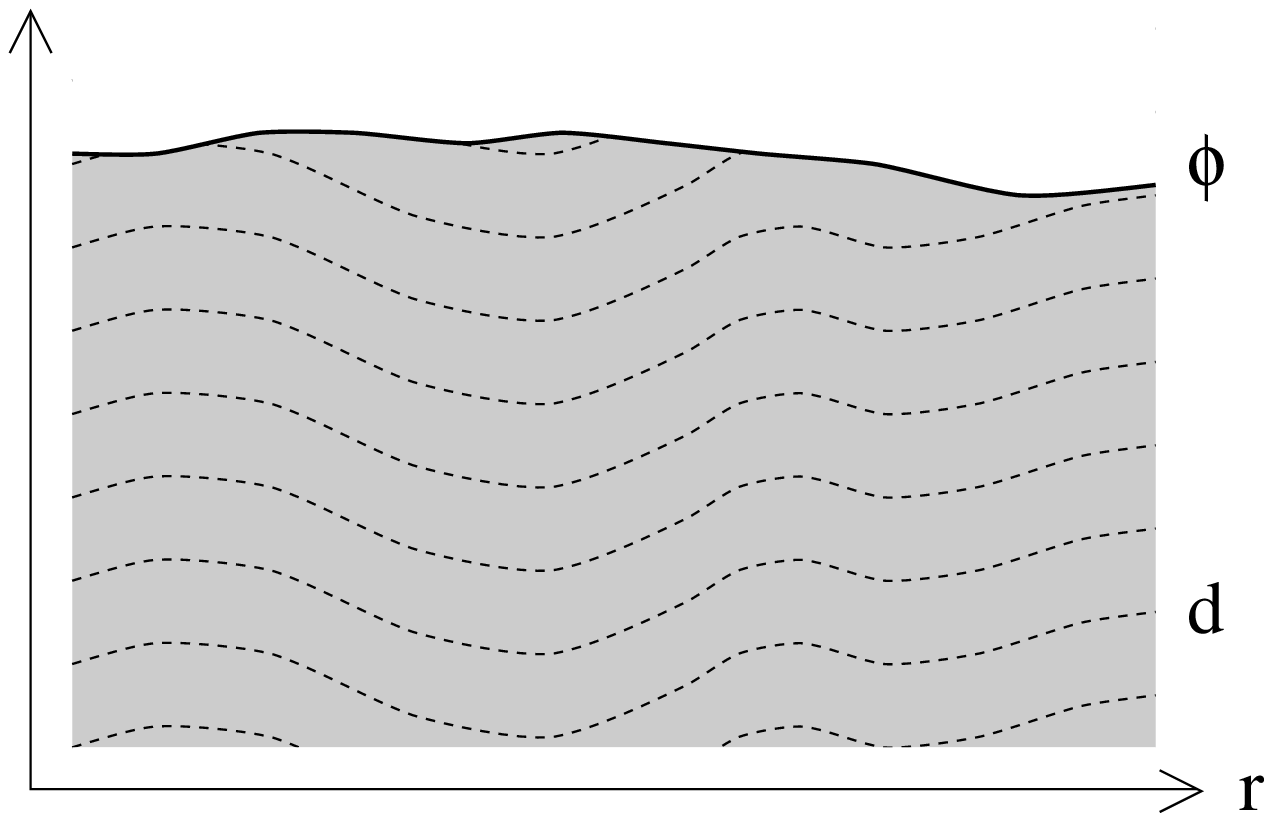}
\caption{
  Schematic sketch of a crystal (hatched area) with interface $\phi$
  (bold line) and disordered crystal planes (dashed lines)
  parameterized by $d$.}
\label{fig.fields}
\end{figure}

The interface profile is described by $\phi$ as a function of time $t$
and the two-dimensional lateral coordinate ${\bf r}$. Its kinetics is
captured by the overdamped equation of motion
\bea{def.eqmo}
m^{-1} \partial_t  \phi(t,{\bf r})&=& K {\bfi \nabla}^2 \phi (t,{\bf r}) -
\chi \sin \left[\phi(t,{\bf r}) - d({\bf r})\right] + \nonumber \\
&&+F +  \xi(t,{\bf r}).
\eea
Here $m$ is a mobility, $K$ is the surface stiffness, $\chi$ is the
amplitude of the periodic pinning potential, which has a quenched
disorder phase $d$ representing deformed crystal planes, $F$ is a
driving force per unit area, and $\xi$ is thermal noise at temperature
$T$. We implicitly suppose a cutoff $\Lambda$ for the wave-vector of
shape fluctuations, which corresponds to an area $4 \pi/\Lambda^2$ per
degree of freedom in the lateral plane. Since we assume regularization
in momentum space, we consider $\phi$ and $d$ to be defined {\em
  continuously} on the whole lateral plane (see Fig.~\ref{fig.fields})
and not just only on a lattice of points ${\bf r}$\cite{note.cont}.

The disorder field is characterized by zero mean and a difference
correlation $\overline { [d ({\bf r})-d({\bf r}')]^2 } = 2 \Delta({\bf
  r}-{\bf r}')$, which for simplicity is taken to be isotropic. For
the following it is convenient to introduce the correlation
($g_0:=\frac 12 m^2 \chi^2$)
\be{def.gamma}
\gamma_0({\bf r}):=g_0 \ \overline{e^{i[d({\bf r})-d({\bf 0})]}}
=g_0 \  e^{-\Delta({\bf r})}.
\ee
In terms of this function, a perfect crystal exhibiting the roughening
transition has $\gamma_0({\bf r})=g$. In the studies of the
superroughening transition \cite{TdV90,RG,TS92}, $\gamma_0({\bf r})$
was supposed to decay rapidly for large ${\bf r}$.
Eq.~(\ref{def.gamma}) shows, that this case corresponds to a very
rough substrate with a rapidly increasing function $\Delta({\bf r})$.
Here we allow for a general, possibly slowly decaying correlation
$\gamma_0({\bf r})$.

In constructing the RG, we choose an approach different from previous
treatments. Since we do not want to specialize to a special
correlation, we have to perform a functional RG. This can be achieved
in the formalism of Martin, Siggia, and Rose \cite{MSR73}, which
requires the introduction of an additional field $\phitil$. In this
formulation the disorder average can easily be performed. The
generating functional then reads
\be{Z}
Z=\int {\cal D}\phitil {\cal D}\phi \ \exp[{\cal A}_0]
\ee
with the action ${\cal A}_0 = {\cal A}^{(0)}_0+ {\cal A}^{(d)}_0 $
composed of free part and disorder part
\bea{def.A}
{\cal A}^{(0)}_0&=&\int dt \ d^2r \
\Big\{\Jtil_0(t,{\bf r}) \phitil_0(t,{\bf r})+
 \frac 12 \vartheta_0 \phitil_0^2(t,{\bf r}) - \nonumber \\
& &- \phitil_0(t,{\bf r})
\left[ \mu_0^{-1} \dot \phi_0(t,{\bf r}) -
\kappa_0 {\bfi \nabla}^2 \phi_0(t,{\bf r}) \right]  \Big\} , \nonumber \\
{\cal A}^{(d)}_0&=&\int dt \ dt' \ d^2r \ d^2r' \
\frac 12 \gamma_0({\bf r}-{\bf r}') \times \nonumber \\
& &\times \phitil_0(t,{\bf r}) \phitil_0(t',{\bf r}')
\cos[\phi_0(t,{\bf r})-\phi_0(t',{\bf r}')].
\eea
We introduced the unrenormalized quantities $\phi_0=\phi$,
$\vartheta_0=2 m T$, $\mu_0=1$, $\kappa_0=mK$, and $\Jtil_0=mF$.

In the absence of the pinning potential ($\chi=0$ and $\gamma_0=0$)
one has correlations
\bea{C^{(0)}}
C^{(0)}(t,{\bf r})&:=& \frac 12
\langle [\phi(t,{\bf r})-\phi(0,{\bf 0})]^2 \rangle^{(0)} \nonumber \\
&=& n \hat C^{(0)}(\mu \kappa \Lambda^2 t,\Lambda r) , \nonumber \\
\hat C^{(0)}(\tau,\rho)&=& 2 \int_0^1 \frac{dx}x
\left[1-{\rm J}_0(x \rho)e^{-x^2 |\tau|}\right] \nonumber\\
&\approx& \ln \max(1,|\tau|,\rho^2) .
\eea
J$_0$ denotes the Bessel function and $n:= \mu \vartheta/(8 \pi
\kappa)=T/(4 \pi K)$ is the roughness coefficient of the unpinned
interface.

We perform a momentum-shell RG \cite{WH73} on model (\ref{Z})
integrating out all modes with wave vectors $\Lambda e^{-dl} < k \leq
\Lambda$ of both fields $\phi$ and $\phitil$. This calculation is
performed only to first order in $\gamma$ without restricting its
functional form. In order to restore the original value of the cutoff,
we then rescale ${\bf r} \to e^{dl} {\bf r}$, and simultaneously $t
\to e^{2 dl} t$, $\phitil_l \to e^{- 2 dl} \phitil_{l+dl}$, and
$\phi_l \to \phi_{l+dl}$ according to the scale invariance in absence
of pinning.

This procedure easily gives the flow equation for the correlator
\be{flow.gamma}
\partial_l \gamma_l({\bf r})=(4+ {\bf r}\cdot {\bfi \nabla}-2n_l)
\gamma_l({\bf r}),
\ee
which is {\em exact} in this order. On the same level, there is no
renormalization of $\mu$, $\vartheta$, and $\kappa$. However, new
interaction terms are generated, which are not present in the
unrenormalized action. In order to keep the theory simple, i.e. the
number of parameters small, we replace these terms {\it approximately}
by contributions to the flow of $\mu$, $\vartheta$, and $\kappa$.
Following Nozi\`eres and Gallet \cite{NG87}, we thereby take the
coupling between the smooth modes (small $k$) into account in order to
avoid artifacts related to the special choice of the cutoff procedure.
The resulting flow equations are (${\bfi \rho}=\Lambda {\bf r}$)
\bea{flow}
\partial_l \kappa_l &=&  \frac {1}{8 \pi \kappa_l \Lambda^4}
\int d^2\rho \ \rho^2 \gamma_l({\bfi \rho}/\Lambda) A(n_l,\rho)  \nonumber\\
\partial_l \vartheta_l &=&
\frac {\vartheta_l}{2 \pi \kappa_l^2 \Lambda^4}
\int d^2\rho \ \gamma_l({\bfi \rho}/\Lambda) B(n_l,\rho)  \nonumber\\
\partial_l \mu_l &=& - \frac { \mu_l}{\vartheta_l}\partial_l \vartheta_l
\nonumber \\
\partial_l \Jtil_l &=& 2 \Jtil_l
\eea
with the definitions
\bea{defs}
n_l&:=& \frac {\mu_l \vartheta_l}{8\pi \kappa_l}  \nonumber\\
A(n,\rho)&:=& {\rm J}_0(\rho) \ e^{-n \hat C^{(0)} (0,\rho)} \nonumber\\
B(n,\rho)&:=& {\rm J}_0(\rho) \int_0^\infty d \tau \
e^{-\tau-n \hat C^{(0)} (\tau,\rho)} .
\eea

Under renormalization the effective stiffness of the interface is
increased, $\partial_l \kappa_l >0$, and the roughness amplitude is
decreased, $\partial_l n_l=- (n_l/\kappa_l) \partial_l \kappa_l<0$.
As long as disorder is renormalized to zero, it will have the effect
to {\em reduce} the equal-time roughness given by $n_{\infty}$, since
$C(t,{\bf r}) \approx n_\infty \hat C^{(0)} ( \mu_\infty \kappa_\infty
\Lambda^2 t, \Lambda r)$. The invariance $\partial_l (\mu_l
\vartheta_l) = 0$ expresses the preservation of the
fluctuation-dissipation theorem under renormalization. This allows a
stepwise solution of the flow equations: first for $\gamma$ and
$\kappa$, since their flow equations are closed. These quantities
completely determine the static properties of the system. Afterwards
one can solve for $\vartheta$ or $\mu$, which affect only the
dynamics. Pinning always reduces mobility, $\partial_l \mu_l<0$, even
if disorder is renormalized to zero on large length scales. The
driving force leads to a drift velocity $\langle \dot \phi \rangle
\approx \mu_{l^*} m F$. In the driven case renormalization has to be
stopped at a value $l^*$ of the flow parameter determined by $mF=e^{-2
  l^*} \kappa_{l^*} \Lambda^2$ since a finite drift velocity washes
out the effect of pinning.

Now we discuss the phase boundary separating the regions, where
disorder changes from being irrelevant to being relevant. The flow
equation for $\gamma$ can directly be integrated:
\be{gamma_l}
\gamma_l({\bf r}) = \exp\left[-2 \int_0^l dl' (n_{l'}-2)\right]
\ \gamma_0(e^l {\bf r}).
\ee
In two extreme cases the relevance of disorder is obvious. The first
case $\gamma_0({\bf r})=g_0$ represents perfect crystal planes. From
eq.~(\ref{gamma_l}) we find the boundary at $n_\infty=2$ which is the
usual roughening transition. In the limit $\chi \to 0$ of a weak
potential the transition temperature is $T_{\R}^{(0)} =8 \pi K$. For
finite $\chi$ the transition temperature $T_{\R}$ will be higher than
$T_{\R}^{(0)}$, since $n_\infty<n_0$. In the second case $\gamma_0
({\bf r})=g_0 \delta(\Lambda {\bf r})$ of extremely deformed crystal
planes we find the boundary at $n_\infty=1$. The superroughening
transition occurs at a temperature $T=T_{\SR} = 4 \pi K$ independently
of disorder strength, since here always $n_\infty=n_0$. This
universality of the transition temperature is special for a infinitely
sharp $\gamma_0 ({\bf r})$. A finite width leads to $\partial_l n_l
<0$ and hence an increased transition temperature $T_{\SR} > 4 \pi K$.
Such a shift has been observed recently in simulations of
superroughening \cite{Cule95}. In a comparison with such data, the
lateral lattice spacing should be taken as width.

The decisive interpolation between theses cases of very broad and very
sharp $\gamma({\bf r})$ is provided when $\gamma_0({\bfi \rho}/\Lambda
)\approx c_0 \rho^{-2\alpha}$ for large $r$ with some amplitude
$c_0>0$ and some exponent $\alpha \geq 0$.  This means, that the
crystal planes themselves have logarithmic roughness. Renormalization
preserves the asymptotic exponent $\alpha$, but the amplitude is
changed: $c_l \propto \exp[-2 {\int_0^l dl'(n_{l'}+\alpha-2)}] c_0$.
Thus we expect pinning to become relevant for the large-scale behavior
of the interface at $n_\infty=2-\alpha$.

\begin{figure}
\epsfxsize=\linewidth
\epsfbox{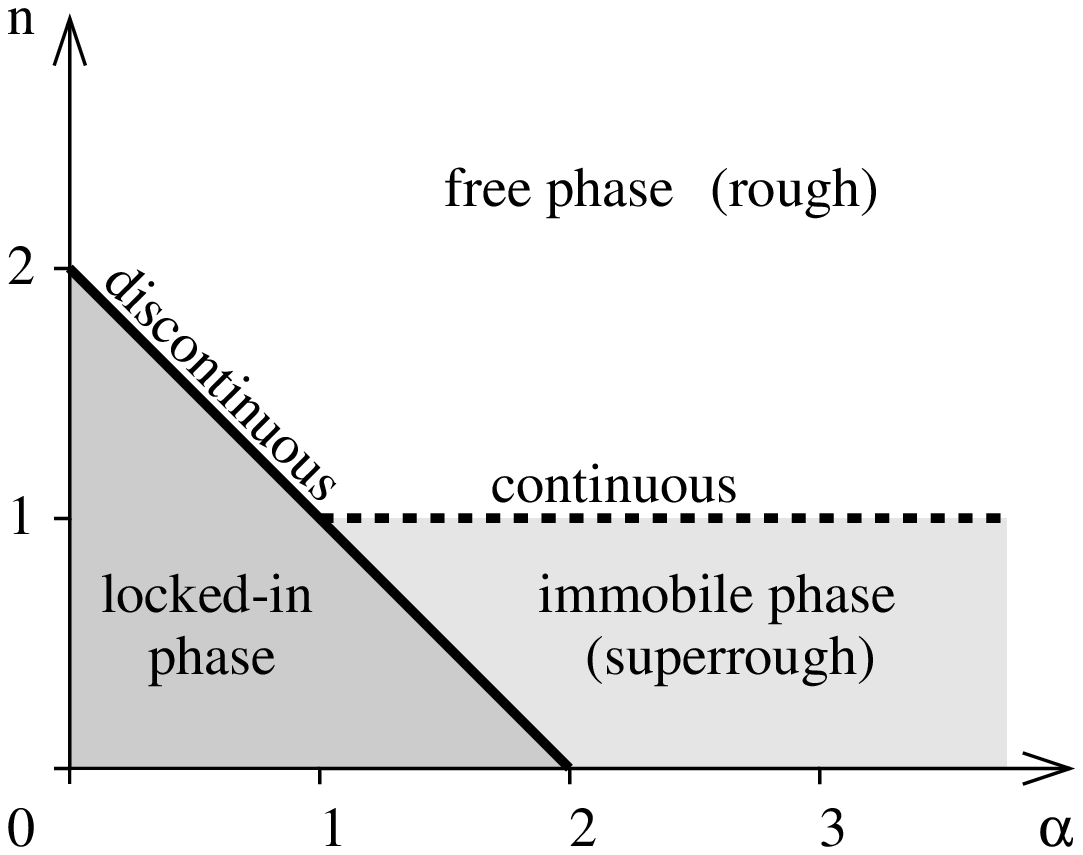}
\caption{
  Phase-diagram: The free phase, where pinning is irrelevant is bound
  by $T/(4 \pi K) \approx n=\max(2-\alpha,1)$. For a range of disorder
  correlations $\alpha \geq 0$ a discontinuous lock-in transition
  occurs, whereas for $\alpha>1$ a continuous superroughening
  transition takes place.}
\label{fig.pd}
\end{figure}

To be more specific, we solve the flow equations (\ref{flow}), after
eq.~(\ref{gamma_l}) has been plugged in. For weak pinning, an
eventual transition is related to large values of the flow parameter,
where the flow equations simplify:
\bea{flow.2}
\partial_l n_l &=&-  y_l^2 \nonumber\\
y_l^2 & \approx &
\left\{
\begin{array}{lll}
A^< e^{-2 \int_0^l dl'(n_{l'}+\alpha-2)} & {\rm for}& \alpha<2 \\
A^> e^{-2 \int_0^l dl'n_{l'}}  & {\rm for}& \alpha>2
\end{array}
\right. \nonumber \\
\partial_l \ln \vartheta_l  & \approx &
\left\{
\begin{array}{lll}
B^<  e^{-2 \int_0^l dl'(n_{l'}+\alpha-2)} & {\rm for}& \alpha<1 \\
B^>  e^{-2 \int_0^l dl'(n_{l'}-1)} & {\rm for}& \alpha>1 .
\end{array}
\right.
\eea
The various coefficients like $A^<$ follow from the original equations
(\ref{flow}) using suitable approximations for large $l$, which are
not displayed here since they are clumsy and straightforward
\cite{note.coeff}. These coefficients are positive and well-defined at
least for $n > 5/4 - \alpha$. This region includes the transitions
under discussion. As long as we are not too close to that border, we
may ignore the implicit dependence of the coefficients on $n$. Due to
the ambivalent asymptotic behavior of flow equations (\ref{flow.2})
one has to distinguish three regimes depending on $\alpha$ displayed
in Fig.~\ref{fig.pd}.

Regime $0 \leq \alpha < 1$. The flow trajectories in the $(n,y)$-plane
have hyperbolic structure $(n_l+\alpha-2)^2-y_l^2=$const. like in the
Kosterlitz-Thouless transition \cite{KT}. In addition,
$\vartheta_l=\vartheta_0 \exp [ - (A^</B^<)(n_l-n_0)]$. Disorder is
irrelevant as long as $n_0+\alpha-2>y_0$ since then $y_\infty=0$ and
$n_\infty>2-\alpha$. For weak disorder strength, the boundary
$n_\infty=2-\alpha$ corresponds to a temperature $T_{\R} (\alpha)
\gsim (2-\alpha) 4 \pi K$. Approaching the transition, $n_\infty$ and
$\vartheta_\infty$ remain finite, as well as the effective mobility
$\mu_\infty=\mu_0 \vartheta_0/\vartheta_\infty$. Thus for $0 \leq
\alpha < 1$ the transition is qualitatively similar to the roughening
case $\alpha=0$. In the low-temperature range, the perturbative flow
equations loose their validity, since the strength of pinning
increases. However, as the effective amplitude $c_l$ of $\gamma_l$
diverges as discussed above, we expect the interface to lock into the
pinning potential.  This implies, that the equal-time roughness is
$C(0,{\bf r}) \sim \alpha \ln (\Lambda^2 r^2)$ below the transition,
whereas it is $C(0,{\bf r}) \sim n_\infty \ln (\Lambda^2 r^2)$ above.
Hence the transition is {\em discontinuous}. Following $\alpha \to
1^-$ along the transition, the magnitude of the jump disappears, since
for $\alpha=n_\infty$ the logarithmic roughness of the crystal planes
and interface coincide. In a similar way the (linear) mobility behaves
discontinuously at the transition since it remains finite approaching
the transition from above and it is zero for the locked-in interface.
The jump of the roughness at the transition from $n_{\infty}=2-\alpha$
to $\alpha$ is {\em universal}, whereas the jump of mobility depends
on all parameters.

Proceeding to the regime $1 < \alpha < 2$, only the flow equation for
$\vartheta$ changes. It exhibits now a continuous divergence, if
$n_\infty$ approaches the value 1 from above, where $T_{\SR} (\alpha)
\gsim 4 \pi K$. Thus the interface becomes immobile ($\mu_\infty=0$)
at that point, although it does not lock in, what happens only at the
lower temperature $T_{\R}(\alpha)$ given before. Thus our flow
equations suggest a sequence of {\em two} transitions: a {\em dynamic}
one and a {\em static} one. However, due to their perturbative nature,
we are not really in position to characterize the low-temperature
phase $n<1$. Possibly the lock-in transition for $1<\alpha <2$ might
be an artifact of the first order calculation. If the transition
exists, the interface is rougher below the transition than above.

In the third regime $2 < \alpha$ we also have to use the second flow
equation of $n$. Nevertheless, the flow remains hyperbolic with
$n_l^2-y_l^2=$const and every $n_0>0$ is renormalized only by a small
amount. Thus disorder is always irrelevant for $n$, a lock-in
transition becomes impossible. The dynamic transition is the same as
in the regime $1 < \alpha < 2$. Again, we can not seriously say
something for $n<1$, except that the linear mobility should continue
to be zero. The controversy mentioned in the introduction is precisely
about this phase, since the commonly used $\gamma({\bf r})$ decaying
faster than any power belong to $\alpha=\infty$. However, there is
agreement that below the superroughening transition the interface is
at least as rough as at the transition. This fact is not captured by
the first order flow equations.

To sum up, we have performed a dynamical RG for a disordered
sine-Gordon model in the presence of weak pinning with arbitrary
disorder correlations. Thereby roughening and superroughening
transition could be merged for a correlation $\gamma_0({\bf r}) \sim
r^{-2\alpha}$. Disorder is irrelevant for the interface at
temperatures $T \gsim \max(2-\alpha,1) 4 \pi K$. In the range $0\leq
\alpha <1$ the transition is of roughening type with a discontinuous
jump of roughness and mobility. The jump of roughness is universal.
This discontinuity decreases, as $\alpha=1$ is approached. For
$\alpha>1$ the transition is of the continuous superroughening type.
The transition temperature is non-universal in all regimes, as long as
disorder correlations have a finite range.

A similar interpolation between transitions has been found in a study
of vortex glasses \cite{HNV93} with an\-iso\-tropic disorder
correlations, which are able to merge the vortex-glass transition
(equivalent to superroughening) and the Bose-glass transition for
$\gamma(x,y)=g \delta (\Lambda x)$. The latter was found at $n=3$, as
is immediately reproduced by our functional flow equation
(\ref{flow.gamma}). In fact, this equation holds also in the
anisotropic case as long as the anisotropy of $\kappa$ can be
neglected.

The author gratefully acknowledges helpful discussions with J. Krug,
T. Nattermann, and L.-H. Tang.

\begin{references}
\bibitem{note.XY} This is the case if the external field has $p$-fold
  symmetry breaking and $p$ is sufficiently large \cite{RG}.
\end{references}
\end{document}